\documentclass[preprint]{aastex631}
\shorttitle{Supernovae Shock Breakout/Emergence Detection}
\shortauthors{Bayless et al.}
\graphicspath{{./}{figures/}}
\begin{document}

\title{Supernovae Shock Breakout/Emergence Detection Predictions for a Wide-Field X-ray Survey}

\author[0000-0003-0660-5900]{Amanda J. Bayless}
\affiliation{Aerospace Corporation, Remote Sensing Department, El Segundo, CA 90245, USA}
\author[0000-0003-2624-0056]{Chris Fryer}
\affiliation{Los Alamos National Laboratory, Los Alamos, New Mexico, 87545, USA}
\affiliation{Physics Department, University of Arizona, Tucson, AZ 85721, USA}
\affiliation{Physics and Astronomy Department, University of New Mexico, Albuquerque, NM 87131, USA}
\author{Peter J. Brown}
\affiliation{Department of Physics \& Astronomy, Texas A\&M University, Mitchell Institute for Fundamental Physics and Astronomy, College Station, TX 77843}
\author{Patrick Young}
\affiliation{School of Earth and Space Exploration, Arizona State University,  411 N Central Ave, Phoenix, AZ 85004, USA}
\author[0000-0002-5499-953X]{Pete Roming}
\affiliation{Southwest Research Institute, San Antonio, TX, 78238, USA}
\affiliation{University of Texas at San Antonio, Physics and Astronomy Department,  UTSA Circle, San Antonio, TX 78249}
\author[0000-0003-4338-1635]{Michael Davis}
\affiliation{Southwest Research Institute, San Antonio, TX, 78238, USA}
\author{Thomas Lechner}
\affiliation{Southwest Research Institute, San Antonio, TX, 78238, USA}
\affiliation{University of Texas at San Antonio, Physics and Astronomy Department,  UTSA Circle, San Antonio, TX 78249}
\author{Samuel Slocum}
\affiliation{Southwest Research Institute, San Antonio, TX, 78238, USA}
\affiliation{University of Texas at San Antonio, Physics and Astronomy Department,  UTSA Circle, San Antonio, TX 78249}
\author{Janie D. Echon}
\affiliation{Department of Physics, Texas State University, San Marcos, TX, 78666, USA}
\author{Cynthia Froning}
\affiliation{Department of Astronomy, University of Texas at Austin, Austin, TX}

%% Note that the \and command from previous versions of AASTeX is now
%% depreciated in this version as it is no longer necessary. AASTeX 
%% automatically takes care of all commas and "and"s between authors names.

%% AASTeX 6.31 has the new \collaboration and \nocollaboration commands to
%% provide the collaboration status of a group of authors. These commands 
%% can be used either before or after the list of corresponding authors. The
%% argument for \collaboration is the collaboration identifier. Authors are
%% encouraged to surround collaboration identifiers with ()s. The 
%% \nocollaboration command takes no argument and exists to indicate that
%% the nearby authors are not part of surrounding collaborations.

%% Mark off the abstract in the ``abstract'' environment. 
\begin{abstract}

There are currently many large-field surveys operational and planned including the powerful Vera C. Rubin Observatory Legacy Survey of Space and Time.  These surveys will increase the number and diversity of transients dramatically.  However, for some transients, like supernovae (SNe), we can gain more understanding by directed observations (e.g. shock breakout, $\gamma$-ray detections) than by simply increasing the sample size.  For example, the initial emission from these transients can be a powerful probe of these explosions. Upcoming ground-based detectors are not ideally suited to observe the initial emission (shock emergence) of these transients.  These observations require a large field-of-view X-ray mission with a UV follow up within the first hour of shock breakout.  The emission in the first one hour to even one day provides strong constraints on the stellar radius and asymmetries in the outer layers of stars, the properties  of the circumstellar medium (e.g. inhomogeneities in the wind for core-collapse SNe, accreting companion in thermonuclear SNe), and the transition region between these two.  This paper describes a simulation for the number of SNe that could be seen by a large field of view lobster eye X-ray and UV observatory.

\end{abstract}

%% Keywords should appear after the \end{abstract} command. 
%% The AAS Journals now uses Unified Astronomy Thesaurus concepts:
%% https://astrothesaurus.org
%% You will be asked to selected these concepts during the submission process
%% but this old "keyword" functionality is maintained in case authors want
%% to include these concepts in their preprints.
%%\keywords{Classical Novae (251) --- Ultraviolet astronomy(1736) --- History of astronomy(1868) --- Interdisciplinary astronomy(804)}

%% From the front matter, we move on to the body of the paper.
%% Sections are demarcated by \section and \subsection, respectively.
%% Observe the use of the LaTeX \label
%% command after the \subsection to give a symbolic KEY to the
%% subsection for cross-referencing in a \ref command.
%% You can use LaTeX's \ref and \label commands to keep track of
%% cross-references to sections, equations, tables, and figures.
%% That way, if you change the order of any elements, LaTeX will
%% automatically renumber them.
%%
%% We recommend that authors also use the natbib \citep
%% and \citet commands to identify citations.  The citations are
%% tied to the reference list via symbolic KEYs. The KEY corresponds
%% to the KEY in the \bibitem in the reference list below. 

\section{Introduction} \label{sec:intro}

Core-collapse SNe (SNe) mark the end-states of massive stars.  These cosmic explosions eject many of the heavy elements into the galaxies and produce some of the most exotic objects in the universe including neutron stars, pulsars, magnetars, and black holes. % and, possibly, quark stars.  
The engine (either the convection-enhanced, neutrino-driven engine~\cite{1994ApJ...435..339H} or a jet-driven engine~\cite{1993ApJ...405..273W}) behind these SNe is equally extraordinary, pushing the limits of our understanding of nuclear, particle, plasma and atomic physics.  \cite{1934PNAS...20..254B} proposed that these explosions were produced by the collapse of the core of a massive star through electron capture down to a compact star composed primarily of neutrons.  For over 50 years, scientists developed increasingly sophisticated models including improved equations of state for dense matter, neutrino physics and general relativity to extract a small fraction of the potential energy released in the collapse to drive a SN explosion \citep{1966ApJ...143..626C,arn80, burrows2021, mezzacappa2020physical, lentz2015, 2007PhR...442...38J, janka2012}.

Observations of SN 1987A provided clues into a key missing piece in our SN models:  convection.  The early gamma-ray emission, and the doppler profiles of both X-ray/gamma-rays [0.01-10MeV~\cite{1988ApJ...329..820P}] and iron lines [infrared, e.g. the 7150 \AA \, multiplet line~\cite{1992MNRAS.255..671S}] all suggested extensive mixing in the SN and/or its progenitor~\citep[for reviews, see][]{2003ApJ...594..390H,2007IJMPD..16..941F}.  Further observations of pulsar “kicks” and SN remnants also suggested strong asymmetries in the SN engine and its progenitor~\citep{2006ApJS..163..335F,2014Natur.506..339G}.  SN 1987A spurred modelers to study multi-dimensional mixing just above the newly formed neutron star, leading them to the current “convection-enhanced” engine that has become the standard paradigm behind normal Type Ib, Ic, and II SN explosions~\citep{1992ApJ...395..642H,1994ApJ...435..339H,janka2012,lentz2015,2018SSRv..214...33B}.  Observations of the $^{44}$Ti distribution in the Cassiopeia A SN remnant further confirmed the current convection-enhanced paradigm engine behind core collapse SNe~\citep{2014Natur.506..339G,2017ApJ...834...19G}.  With our greater understanding of the SN engine, astronomers were able to make a host of predictions of explosion properties, compact remnant formation (masses and kicks), and nucleosynthetic yields which, in turn, have turned the study of SN into probes of the physics of the universe~\citep[e.g.][]{2001ApJ...554..548F,2012ApJ...749...91F}.   

However, these improved models have their limitations, in particular, in our understanding of the massive stellar progenitors of these explosions.  For massive stars, mass loss from winds, wave- and opacity-driven instabilities, and explosive burning can all drive mass-loss~\citep[for reviews, see][]{2016ApJ...817...66K,2020ApJ...898..123F,2021arXiv211001565L}.  

 Wind-driven mass loss is believed to be metallicity-dependent \citep{2000A&A...360..227N,2002ApJ...577..389K} with a rough mass-loss scaling increasing with the square root of the metallicity.  At high metallicity, this mass loss can remove the hydrogen envelope leading to a Wolf-Rayet phase that can dramatically alter the mass of the stellar core.  The fate of the collapsing star depends sensitively on the mass of the core.  At high metallicities, wind-driven mass loss can alter the fate of a collapsing star from no explosion and black hole formation to a strong supernova explosion and the formation of a neutron star \citep{heg03}.  Understanding this mass-loss is important in understanding the fates of high-metallicity stars.  Another key uncertainty is the extent of mixing in a star.  This mixing determines both the size and angular momentum of the core at collapse~\citep{2018MNRAS.481.2918M,2019MNRAS.484.4645C,2020MNRAS.496.1967K}.  The convection in the silicon shell provides the seeds for convection in the SN engine~\citep{2007ApJ...659.1438F,2020ApJ...901...33F,2022ApJ...924L..15F}.  By constraining these uncertainties in stellar models, we can improve our models for SNe, making them more powerful probes of the multitude of SN applications.

Probing the nature behind stellar mass loss and stellar mixing is difficult and many measurements require population studies (e.g. galactic chemical evolution, compact remnant mass distributions).  Pulsar and X-ray binaries, coupled with merging compact binaries detected in gravitational waves, are beginning to characterize the mass distribution of the compact remnants from this collapse~\citep{2004tsra.conf...42L,2021ApJ...914L..18D}, but all of these measurements must be filtered through binary evolution models to extract the true distribution formed in stellar collapse~\citep{2012ApJ...757...91B}.  Nucleosynthetic yields from individual SNe and SN remnants can place some constraints on the burning layers within a star but these measurements are limited both in the number and accuracy of elemental abundances.  The ejecta from SN light curves or remnants are also altered by the circumstellar medium and asymmetries in that medium~\citep{1982ApJ...258..790C,1996ApJ...472..257B}.

Shock breakout emission is typically used to describe the photons arising when the forward shock of the SN blastwave break outs of the stellar photosphere~\citep{nak10,2011AstL...37..194B,2017hsn..book..967W,2021arXiv210913259I}.  This forward shock accelerates down the steep density gradient in the transition between the stellar edge and the stellar wind, reaching high velocities.  Shock-heating with this blast wave can produce emission in the X-rays and, in some highly-relativistic cases, gamma-rays.  In this simple model, shock breakout can accurately measure the stellar radii.  SN 2008D~\citep{sod08, 10.1086/591522}, a serendipitous discovery of the shock breakout was used to both calibrate these analytic models and calculate the stellar radius of this event.  Other possible observations exist, but most of these few cases are SNe associated with gamma-ray bursts \citep[e.g.][]{cam06} where both the explosion mechanism, shock heating and progenitors are believed to be very different from standard SNe. In addition to the shock breakout mechanism, other mechanisms have been proposed including the breakout of a dense CSM \citep{svi12} and a mildly relativistic jet \citep{10.1126/science.1158088, 10.1063/1.3027928, xu_2008}. Other observations in the optical covering the time of explosion (e.g. high cadence staring by the Kepler spacecraft; \citealp{gar16}) are only sensitive to the Rayleigh-Jeans tail of the thermal emission, missing most of the energy output and lacking the power to constrain the high temperatures \citep{10.1038/nature25151}.
%also Armstrong et al. 2021

 The analytic models described above focus on the energy in the forward shock assuming a spherically symmetric explosion and do not capture the rich physics involved in this early-time emission (e.g.~\cite{nak10}). Asymmetries in the supernova blastwave, stellar envelope, its wind and the transition region between these two regions both alter the breakout timescales~\cite[e.g.][]{2021arXiv210913259I} and produce a myriad of oblique shocks~\cite[e.g.][]{2020ApJ...898..123F, Sal2014, Mat2013}.   The asymmetries drastically change the spherically symmetric picture with its forward and reverse shock dominating the shock heating.  Asymmetries in the shock and inhomogeneities in the wind cause a myriad of shocks traversing the ejecta, causing it to be shocked multiple times. Detailed models of the heating from these multiple shocks can allow us to probe these asymmetries.  However,  detailed models also predict a diverse set of signals (with different durations, peak brightnesses and hardness ratios that can range by over a magnitude) and to probe the physics behind this emission, we must characterize the full distribution of the early emission.  Although this emission is far broader than the what arises from shock breakout, for the rest of this paper, we will refer to it as shock breakout emission.

 Since the shock breakout phase is on the order of 100-1000 seconds, detecting this early-time phase is inherently difficult because it requires high-energy instruments surveying a large field-of-view ($\approx0.2$ sr).
In this paper, we 
%review the required instruments needed
introduce the necessary instruments and the required key performance characteristics needed for such a survey (Section~\ref{sec:detector}).  We present a first step in more detailed models of this early-time emission calibrated with existing X-ray and UV observations (Section~\ref{sec:snemission}).  The broad range of signals produced by the asymmetries in the explosion can be studied, but requires a large sample of observations.  We discuss the sample requirements in Section~\ref{sec:sample}.  We combine the theoretical emission models with SN rates and the detector characteristics to calculate detection rates in Section~\ref{sec:rate}.  We conclude with a discussion of the scientific studies the combined X-ray/UV mission could perform on these detected early-time SN transients.

\section{Instrument Requirements for Shock Breakout Detection}
\label{sec:detector}

%To detect and localize shock breakout events a wide-field, soft X-ray instrument is required. The wide field is required in order to have a large enough volume of space to detect a sufficient number of SNe, while the soft X-ray sensitivity is required in order to capture the peak of the shock breakout emission. For our simulations, we assume an X-ray instrument with a 0.3 sr field-of-view and a 5$\sigma$-sensitivity in a 24 s exposure in the 0.1-5\,keV band of $1.2\times10^{-10} {\rm erg ~cm^{-2} s^{-1}}$.

%For follow-up of the cooler post-shock break out photons, a far-UV instrument is required. For our simulations, we assume a telescope sensitive in the 1350-3000 \AA\ wavelength region, with a 5$\sigma$-sensitivity in a 600 s exposure of $1.4 \times 10^{-14} {\rm erg \, cm^{-2} nm^{-1} s^{-1}}$.

To detect and localize shock breakout events from Wolf-Rayets (WR), Red Supergiants (RSG), and Blue Supergiants (BSG) a wide-field, soft X-ray instrument is required. The X-ray instrument’s wide field is required in order to observe a large enough volume of space to detect a sufficient number (see section~\ref{sec:sample}) of shock breakout events, while the soft X-ray sensitivity is required in order to capture the peak of the shock breakout emission. For follow-up of the cooler post-shock breakout photons, a far-UV instrument is required. The key instrument performance characteristics required for achieving the science objectives outlined are provided in Table~\ref{tab:Instrument}.

\begin{table*}[htb]
\footnotesize
\center
\caption{Key Instrument Performance Characteristics}
\label{tab:Instrument}
\begin{tabular}{lcc}
\hline
Characteristic & X-ray Instrument  & Far-UV Instrument\\
\hline
\hline

Spectral Range          & 0.1-10 keV                            & 136-300 nm        \\
%Sensitivity ($5\sigma$) & $1.2\times10^{-10}$ erg/s/cm$^{2}$    & $1.0\times10^{-16}$ erg/s/cm$^{2}$ \\ 
%                        & (24-s exposure)                       & (1600-s exposure) \\
Sensitivity ($5\sigma$) & $1.2\times10^{-10}$ erg/s/cm$^{2}$    & $8.0\times10^{-14}$ erg/s/cm$^{2}$ \\ 
                        & (24-s exposure)                       & (24-s exposure) \\                       
Field-of-view           & 0.20 sr                               & 900 sq-arcmin     \\
Temporal Cadence        & 8 s                                   & 0.3 s             \\
Localization Accuracy   & 3.1$'$                                & 1.0$''$            \\

\hline
\end{tabular}
\end{table*}

\section{SN Emission Models}
\label{sec:snemission}

To calculate the detection rate of SNe, we must first develop models for their shock-heated X-ray emission.  We include both analytic models of traditional shock breakout (emission limited to the energy in the forward shock~\cite{nak10}) and simulations that include both shock heating from shocks in the stellar envelope, in the stellar wind, and in the transition region between the edge of the star and the wind.   For both the analytic and simulated models, we constrain the theoretical uncertainties using observations of shock breakout and, in the case of the simulated models, the later UV emission.  For the analytic models, we use the peak luminosities predicted by the analytic models, but set the the variation in the peak fluxes ($1-\sigma$) variation in the peak fluxes from SN 2006aj and SN 2008D.  For the simulated models, we use the post-breakout UV observations to help calibrate the models beyond the few shock breakout observations (see discussion in Section~\ref{sec:simulations}).  As we shall see, the dearth of prompt emission observations allows for a broad range of models and the variations in these models are a primary source of error in our detection rate. These models predict peak luminosities along with e-folding duration times for the shock emission that, when combined with the rates of different types of supernovae (discussed in  Section \ref{sec:rate}), can be used to calculate the detection rate of SNe for a mission like that described in Section \ref{sec:detector}.  In this section, we describe both the analytic and simulation models used to estimate this X-ray emission.  Because UV observations help to calibrate our models, we include a discussion of the UV emission as well.

\subsection{Analytic Models of Shock Breakout}

Shock breakout emission depends upon a broad set of stellar and explosion properties and, as such, can probe these properties.  First and foremost is our understanding of the stellar radius and the transition region between the edge of the star and its wind.  In compact binaries, understanding the edge of the star is important in dictating mass transfer and common envelope events~\citep{1999ApJ...526..152F}.  Understanding the transition region is also important in understanding the nature of stellar winds.  Little is known about the density conditions in the transition region between the edge of a massive star and its stellar wind.  It is expected that there is a transition region where the density drops from typical stellar envelope densities to the much lower wind densities, but the exact nature of the density profile is not well understood.  Most of the work studying early X-rays focuses on shock-heating from the material in the SN blast-wave that accelerates down the steep density gradient.  The shock acceleration can be understood in the non-relativistic limit using the Taylor-von Neuman - Sedov similarity solution~\citep{1977MoIzN....Q....S} using unit analyses (e.g. $[E_{\rm exp}]/[\rho] = [r]^5/[t]^2$).  If we assume $\rho=\rho_0 r^{-\gamma}$ and the velocity (v) is determined by the time derivative of $r$, we :
\begin{equation}
    v \propto (E/\rho_0)^{1/(5-\gamma)} t^{(\gamma-3)/(5-\gamma)}.
\end{equation}
If the density gradient is steep (i.e. $\gamma=4$), the forward shock accelerates.  This acceleration can lead to relativistic velocities~\citep{1974ApJ...187..333C,2001ApJ...551..946T} and this shock acceleration model formed the first theoretical engine for gamma-ray bursts~\citep{1974ApJ...187..333C}.

Although it is difficult to achieve the extremely high velocities to produce gamma-ray bursts via Colgate's shock acceleration mechanism with normal supernova explosions, the shock acceleration can produce sufficiently high-velocity material such that its shock heating produces thermal emission in the X-ray band.  For this shock breakout emission, we use standard analytic models based on ~\citet{nak10} with a Gaussian shaped light curve in log luminosity -- log time space.  We calibrate the width and peak luminosities of the Gaussian models on the two existing shock breakout observations:  SN 2008D \citep{mod09} and SN 2006aj/GRB060218 \citep{cam06}.  These two observed X-ray light curves can  also be modeled as Gaussians in log luminosity -- log time space. The maximum X-ray luminosity for SN 2008D is $Log [L_{\rm X-ray} /({\rm  erg\,  s^{-1}})] = 43.6$ and the light curve has a FWHM of 250 sec. For SN 2006aj/GRB060218, the maximum X-ray luminosity is $Log [L_{\rm X-ray}/({\rm  erg \, s^{-1}})] = 45.4$ and has a FWHM of 14,954 sec.  Using these two data points we interpolate a FWHM for each of the simulated Gaussian light curves' assigned maximum luminosity. The more luminous the peak in the simulation, the broader the light curve will be. 

The analytic solutions of~\citet{nak10}  predict an X-ray and UV flux, but not the variation in the flux.  Using SN 2006aj and SN 2008D as a guide to establish a $\approx 1-\sigma$ variation, we use these analytic models and their variation in total flux and duration to assign an average maximum peak X-ray luminosity in erg/s for each SN based on type:  $Log [L_{\rm X-ray}/( \, {\rm  erg \, s^{-1}})] = 43, 44, 45, 44$ for red supergiants (RSG), blue supergiants (BSG), Wolf-Rayet stars (WR), and other/peculiar systems respectively. These peak values are from the results of \citet{nak10}. For the simulation, we wanted to vary the luminosity of a particular event.  We use a FWHM Gaussian average distribution of $Log (L_{\rm X-ray})$, with $\sigma = 1$  for all types.  The luminosity for a single SN event in the simulation is a random value chosen within the distribution. We also tested a lower maximum luminosity based on the UV LANL simulations (see Section 3.2) for RSG ($Log [L_{\rm X-ray} /({\rm  erg\,  s^{-1}})]  = 42$) and WR ($Log [L_{\rm X-ray} /({\rm  erg\,  s^{-1}})] =43$), but with a wider ($\sigma = 3$) average distribution (see Section 5.2). 
%(Figure \ref{lxhist}).  
%Thus, each SN event in the simulation receives a unique maximum $Log (L_{\rm X-ray})$ from a random number generator based on this average distribution.  The distribution of peak luminosities for these models are shown in Figure~\ref{lxhist} and 
The properties of the models for the simulation are detailed in Table~\ref{tab:Models}.

%\begin{figure}[htb]
%\center 
%\includegraphics[scale=0.34,angle=0]{LxHist.eps}
%\caption{Distribution of maximum X-ray luminosity for each SN type in the simulation.}
%\label{lxhist}
%\end{figure}

\begin{table*}[htb]
\footnotesize
\center
\caption{Light-Curve Models}
\label{tab:Models}
\begin{tabular}{lcccccc}
\hline
Model & $L^p_{>0.1keV}$  & $t^{dur}_{> 0.1keV}$ & $L^p_{>0.3keV}$ & $t^{dur}_{> 0.1keV}$ & $L^p_{UV}$ & $t^{dur}_{UV}$\\
 & (Log ${\rm erg s^{-1}}$) & (s) & (Log ${\rm erg s^{-1}}$) & (s) & (Log ${\rm erg s^{-1}}$) & (s) \\
\hline
\hline
Analytic & & & & & & \\
\hline
RSG$_{\rm Analytic}$ & 43 &  & 43 & & & \\
BSG$_{\rm Analytic}$ & 44 & & 44 & & & \\
WR$_{\rm Analytic}$ & 45 & & 45 & & & \\
\hline
\hline
Simulations & & & & & & \\
\hline
RSG1 & 45.2 & 600 & 42.1 & 5000 & 43.9 & 400 \\
RSG2 & 45.5 & 1000 & 43.0 & 600 & 43.7 & 3000 \\
RSG3 & 43.0 & 500 & 36.0 & 3000 & 42.2 & 10,000 \\
RSG4 & 43.2 & 450 & 35.4 & 3000 & 42.1 & 20,000 \\
RSG5 & 43.2 & 450 & 35.4 & 3000 & 42.1 & 20,000 \\
RSG6 & 45.3 & 1600 & 42.7 & 600 & 43.9 & 2000 \\
RSG7 & 45.2 & 800 & 42.4 & 600 & 43.8 & 400 \\
RSG8 & 45.2 & 600 & 42.1 & 6000 & 43.9 & 400 \\
RSG9 & 45.5 & 1000 & 43.2 & 600 & 43.3 & 3000 \\
\hline
WR-A series & & & & & & \\
WRA1 & 42.6 & 1000 & 42.0 & 3000 & 42.2 & 10000 \\
WRA2 & 42.5 & 3000 & 41.3 & 3000 & 42.0 & 10000 \\
WRA3 & 42.3 & 400 & 41.2 & 600 & 42.1 & 1000 \\
WRA4 & 44.6 & 500 & 43.5 & 3000 & 43.2 & 700 \\
WRA5 & 43.4 & 700 & 41.6 & 300 & 42.5 & 400 \\
WRA6 & 48.9 & 200 & 49.0 & 200 & 44.7 & 200 \\
WRA7 & 45.5 & 200 & 44.5 & 200 & 42.7 & 200 \\
WRA8 & 41.1 & 100 & 40.2 & 100 & 40.6 & 200 \\
WRA9 & 43.2 & 100 & 42.1 & 100 & 40.7 & 100 \\
\hline
WR-B series & & & & & & \\
WRB1 & 46.0 & 10,000 & 44.9 & 5000 & 43.3 & 30,000 \\
WRB2 & 45.8 & 2500 & 44.6 & 7000 & 43.5 & 25,000 \\
WRB3 & 45.7 & 1600 & 44.2 & 10,000 & 43.0 & 20,000 \\
WRB4 & 45.7 & 1600 & 44.2 & 10,000 & 43.0 & 20,000 \\
WRB5 & 45.5 & 20,000 & 43.6 & 10,000 & 43.2 & 30,000 \\
WRB6 & 46.5 & 30,000 & 46.0 & 5000 & 44.4 & 80,000 \\
WRB7 & 46.1 & 30,000 & 44.9 & 10000 & 43.9 & 10,000 \\
WRB8 & 46.5 & 20,000 & 46.0 & 6000 & 44.5 & 60,000 \\
WRB9 & 46.2 & 35,000 & 46.0 & 5000 & 44.5 & 70,000 \\
\hline
WR-C series & & & & & & \\
WRC1 & 39.7 & 1500 & 31.9 & 9000 & 41.1 & 80,000 \\
WRC2 & 39.3 & 2000 & 31.9 & 30,000 & 40.0 & 100,000 \\
WRC3 & 38.5 & 5000 & 32.0 & 50,000 & 40.1 & 80,000 \\
WRC4 & 42.0 & 2000 & 39.6 & 1500 & 42.1 & 10,000 \\
WRC5 & 41.1 & 3000 & 38.2 & 1500 & 41.1 & 15,000 \\
WRC6 & 41.2 & 2000 & 37.2 & 1500 & 41.5 & 5000 \\
WRC7 & 40.3 & 5000 & 35.2 & 2500 & 41.9 & 7000 \\
WRC8 & 39.8 & 1000 & 31.8 & 5000 & 40.9 & 10,000 \\
WRC9 & 36.0 & 2000 & 30.7 & 10,000 & 40.3 & 20,000 \\
\hline
\end{tabular}
\tablenotetext{}{\footnotesize In the RSG runs, the different models correspond to different ejecta masses and shock heating (due to variations in the inhomogeneities in the star).  Early-time UV cooling fluxes suggest a wide variation in WR models and we created 3 sets of extreme shock heating model suites for WR stars:  WRA, WRB, WRC.  Within those models, we also vary the ejecta masses and density profiles.}
\end{table*}

\begin{figure*}[htb]
\includegraphics[scale=0.4]{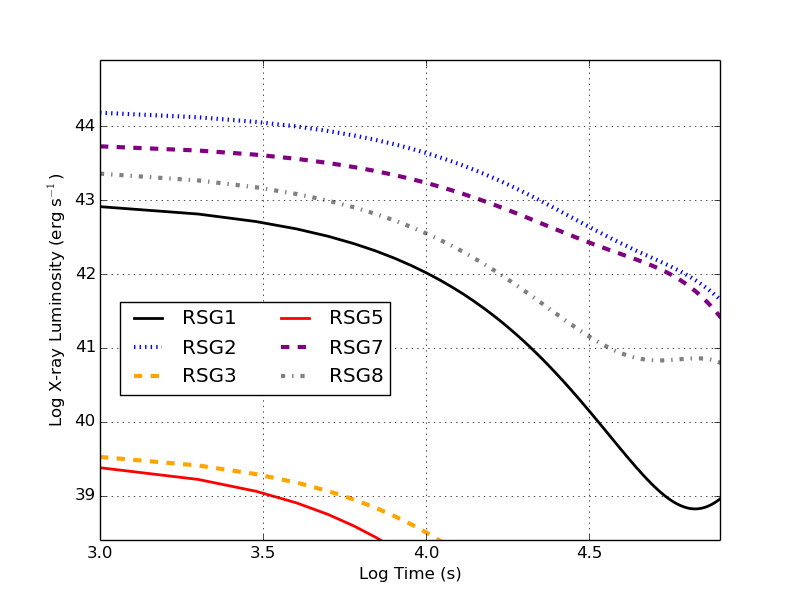}
\includegraphics[scale=0.4]{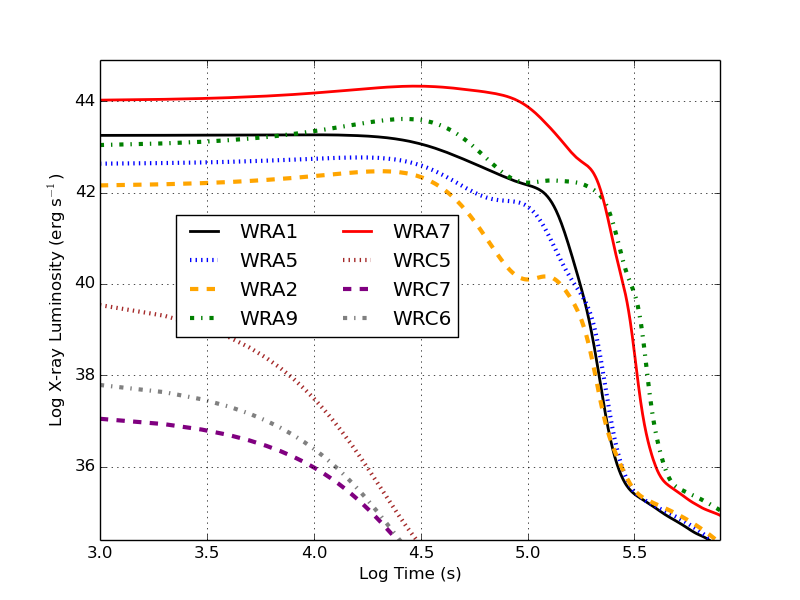}
\caption{X-ray Luminosity ($>2$keV) of a subsample of our simulated SBO lightcurves for RSG (left) and WR stars (right).  The models producing these light-curves vary the stellar radii, explosion energy and shock heating in the star and stellar wind, see Section~\ref{sec:simulations}.}
\label{fryermodels}
\end{figure*}

For the LANL models, we assigned a maximum UV luminosity of $log (L_{UV}) = log (L_{X-ray}) - 0.5$ (from our models in Table~\ref{tab:Models}, this value can vary by several magnitudes, but the average is $\sim 0.5$).  This is based on LANL simulations of spherically-symmetric shock breakout following the assumptions in the \citet{nak10} X-ray models.  Figure \ref{fryermodels} shows a sample of the X-ray light curves from this simulation.  However, as we see in Table~\ref{tab:Models}, the peak UV emission can vary considerably with respect to the X-ray (see Section~\ref{sec:simulations}) and this assumption for our analytic models is approximate at best.

\subsection{Calibrated Simulations}
\label{sec:simulations}

Analytic models make a series of simplifying approximations that can lead to errors in the light-curve profile.  For example, most analytic models do not include asymmetries in the explosion and, incorporating this effect can dramatically lengthen the duration of the shock breakout signal~\citep{2021arXiv210913259I}.  These analytic models also do not include shock heating and, even in 1-dimensional models, this shock heating can alter the temperature profile and change the light-curve evolution~\citep{bay15}.  The oblique shocks produced in asymmetric explosions and the propagation of shocks through inhomogeneous media (stellar envelope or wind) can exacerbate this effect.  In this section, we develop a set of simulated light-curves that include the full temperature profile of the expanding supernova blastwave and a prescription for shock heating.

Our simulated models are based on a simulation database using the LANL light-curve code that models SN light-curves assuming homologous-outflows and using diffusive radiation-transport~\citep{delarosa17}.  For the LANL models, we use a single gray opacity assuming electron scattering dominates the opacity for the transport.  We post-process these simulations using LTE opacities assuming solar-metallicity abundances to produce spectra and band light-curves\footnote{At shock breakout, radiation and electrons can decouple and using a single temperature to describe both of these distributions in calculating the opacities becomes an increasingly bad assumption, especially for line strengths.  As we are focused on broadband signatures where electron scattering can play a major role (which is not so sensitive to the break with LTE), the effect of assuming LTE is minimized.  Given the many other uncertainties, shock heating, stellar profiles, etc., we believe this is not a major uncertainty in this study.}. This code is designed to allow a wide range of parameterizations for  the initial conditions: mass, radius, kinetic energy of the explosion, shock heating that determines the temperature profile (although this should scale with the kinetic energy, asphericities in the stellar envelope and SN shock-produced secondary shocks that will alter this result), density profile (assuming a power law or 2-component power law), ejecta mass, and $^{56}$Ni mass.   Shock heating can occur in the forward shock alone, causing heating in a narrow shell, but it is more likely that the shock heating is more widespread as the asymmetric shock plows through the heterogeneous medium of the stellar envelope and stellar wind.  Our implementation allows a parameterized shock temperature (${\rm Temp_{\rm shock}}$) that can be placed in a shell or a larger part of the star (see Table~\ref{tab:param}).

One of the biggest uncertainties in these calculations is the implementation of the shock heating.  Much more work needs to be done to understand the extent of shock heating in supernova explosions, including both detailed radiation-hydrodynamics models in multi-dimensions.  In this study, we instead parameterize this shock heating, calibrating not only to shock-breakout observations, but cooling observations in the UV.

For our base models, we use stellar radii of $5\times10^{10}$, $1\times10^{12}$, and $8\times10^{13}~{\rm cm}$ for WR, BSG and RSG respectively.  Because of internal shocks, the ejecta can be hot and the masses of the stars can drastically alter the duration of the emission, particularly the UV.  Although we vary the ejecta masses, our basic models assume $4\,M_\odot$ of ejecta mass for the Wolf-Rayet star (this corresponds to a $5.5\,M_\odot$ core mass at collapse or the helium core mass of a star with a zero-age main sequence mass of $20-25\,M_\odot$) and $12\,M_\odot$ of ejecta (corresponding to a $13.5\,M_\odot$ stellar mass at collapse) for the blue and red supergiants.   Based on current 1-dimensional stellar models, it is believed that the density profile  can be fit by a power-law ($\rho \propto r^{-\alpha}$) and, especially at the edge of these stars, this power-law can be quite steep.  Our nominal value for $\alpha$ at this edge is 6 based on studying pre-collapse progenitors from MESA from \cite{fryer18}. However, this power-law depends on the prescription for the outer boundary of these models and the exact value is uncertain and this effects the temperature of the shock.  For our simulated models, we vary a range of explosion properties including progenitor mass, explosion energy, the temperature profile caused by shocks (including some models where only a shell of material is heated), the nickel yield, and the density profile (Table~\ref{tab:param}).  Table~\ref{tab:Models} shows a range of results for different shock breakout emission varying ejecta mass, shock heating and density profiles in the envelope, transition region and wind (for the Wolf-Rayet Stars).

\begin{table*}[htb]
\footnotesize
\center
\caption{Model Light Curve Parameters}
\label{tab:param}
\begin{tabular}{lccccc}
\hline
Model & M$_{\rm exp}$  & E$_{\rm exp}$ & R$_{\rm star}$ & Temp$_{\rm shock}$ & Density \\
 & (M$_\odot$) & ($10^{51}$erg) & (R$_\odot$) & (keV) & $\alpha$ ($\rho \propto r^{-\alpha}$)\\
\hline
\hline
RSG1 & 15.0 & 1.0 & 11.0 & 0.04 & 6 \\
RSG2 & 8.0 & 1.0 & 11.0 & 0.05 & 6 \\
RSG3 & 8.0 & 1.0 & 11.0 & 0.02 & 6 \\
RSG4 & 10.0 & 1.0 & 11.0 & 0.018 & 6 \\
RSG5 & 8.0 & 1.0 & 11.0 & 0.02 ($M_{\rm Ni}=0.1M_\odot$) & 6 \\
RSG4 & 10.0 & 1.0 & 11.0 & 0.018 & 6 \\
RSG7 & 12.0 & 1.0 & 11.0 & 0.016 & 6 \\
RSG8 & 12.0 & 1.0 & 11.0 & 0.016 ($M_{\rm Ni}=0.1M_\odot$) & 6 \\
RSG9 & 8.0 & 1.0 & 11.0 & 0.05 ($M_{\rm Ni}=0.1M_\odot$) & 6 \\
\hline
WRA1 & 3.0 & 1.0 & 0.8 & 2.7 & 6 \\
WRA2 & 3.0 & 1.0 & 0.8 & 2.4 & 6 \\
WRA3 & 1.0 & 1.0 & 1.0 & 1.1 & 6 \\
WRA4 & 1.0 & 1.0 & 1.0 & 2.4 & 6 \\
WRA5 & 1.0 & 1.0 & 1.0 & 10.0 (shell) & 6 \\
WRA6 & 1.0 & 1.0 & 1.0 & 100.0 (shell) & 9 \\
WRA7 & 1.0 & 1.0 & 1.0 & 15.0 (shell) & 9 \\
WRA8 & 1.0 & 1.0 & 1.0 & 3.8 (shell) & 9  \\
WRA9 & 1.0 & 1.0 & 1.0 & 5.5 (shell) & 9 \\
\hline
WRB1 & 5.0 & 1.0 & 0.8 & 9.3 & 6 \\
WRB2 & 3.0 & 1.0 & 0.8 & 10.0 & 6 \\
WRB3 & 8.0 & 1.0 & 0.8 & 8.9 & 6 \\
WRB4 & 5.0 & 1.0 & 0.8 & 8.5 & 6 \\
WRB5 & 8.0 & 1.0 & 0.8 & 7.9 & 6 \\
WRB6 & 4.0 & 1.0 & 0.8 & 10.5 & 6 \\
WRB7 & 8.0 & 1.0 & 0.8 & 9.5 & 6 \\
WRB8 & 3.0 & 1.0 & 0.8 & 11.0 & 6 \\
WRB9 & 5.0 & 1.0 & 0.8 & 10.4 & 6 \\
\hline
WRC1 & 3.0 & 1.0 & 0.8 & 0.93 & 6 \\
WRC2 & 5.0 & 1.0 & 0.8 & 0.88 & 6 \\
WRC3 & 3.0 & 0.5 & 0.8 & 0.86 & 6 \\
WRC4 & 3.0 & 1.0 & 0.8 & 1.8 & 6 \\
WRC5 & 5.0 & 1.0 & 0.8 & 1.75 & 6 \\
WRC6 & 3.0 & 0.5 & 0.8 & 1.73 & 6 \\
WRC7 & 8.0 & 1.0 & 0.8 & 1.8 & 6 \\
WRC8 & 8.0 & 0.5 & 0.8 & 0.85 & 6 \\
WRC9 & 5.0 & 0.1 & 0.8 & 0.30 & 6 \\

\hline
\hline
\end{tabular}
\begin{tablenotetext}
\footnotesize
\item $M_{\rm Ni}=0.2M_\odot$ models increase $^{56}$Ni mass by a factor of 10 from standard calculations to determine its effect (small on shock breakout).  WRA models studied the effects of heating limited to the forward shock (no reverse or multiple shocks as is expected in the more heterogeneous stars/winds).  WRB/WRC models focus on the dependence of shock temperature and progenitor mass.  For those models labeled "shell", we restrict shock heating to a narrow outer shell of the blastwave.  
\end{tablenotetext}
\end{table*}

Early time X-ray emission occurs through a number of shock-interaction sources:  shocks produced as the ejecta propagates through convective asymmetries in red supergiant envelopes, shocks produced by the ejecta front accelerated as the SN blast propagates from the stellar edge to its wind (spherically-symmetric shock breakout from our analytic models), shocks produced as the ejecta propagates through an asymmetric stellar wind around a Wolf-Rayet star.  This breakout emission is short-lived and, as we discussed above in our analytic models, is only constrained by two observed systems:  SN 2008D and SN2006aj.  Our models suggest that the distribution of luminosities is much broader than these two observations suggest and our models extend both brighter and dimmer than these two observations.  Modeling the signal from traditional shock breakout requires resolving the thin layer of material that accelerates to high velocities as the shock propagates through the poorly understood star/wind transition region.  Our theoretical understanding of these uncertainties is inadequate, and this emission has been argued to produce both GRBs and shock breakout signals alike.  Because this emission requires extremely high spatial resolution of the edge of the star, we only capture this emission in  a subset of highly-resolved simulations (WR-A series).    The X-ray signal above 0.3\,keV can be considerably dimmer than the signal above 0.1\,keV and the sensitivity below 0.3\,keV is important in future X-ray detectors.

Shocks produced as the SN blast waves propagates through the convective red supergiant envelope or the Wolf Rayet winds depend upon the magnitude and structure of the asymmetries.   We have constructed two series of models corresponding to bright (WR-B) or dim (WR-C) extremes.  These WR models can have much longer X-ray outbursts than traditional shock breakout models.  However, the duration of these X-rays are not so long that we can use the wealth of late-time X-rays observed in core-collapse SNe (typically believed to be produced by shock interactions with shells of material in the circumstellar medium).  The brightest models (WR-B) can be as bright or brighter than both shock breakout and current observations.  In Section~\ref{sec:rate}, we show that for many of our models, some prompt emission can be observed out to 200\,Mpc with current detector technology.  However, this will not be the case for our dimmer models with less shock heating (WR-C suite).  A sample of the models used in this study are listed in Table~\ref{tab:Models}. Although the X-ray constraints are limited, these long-term blast models predict even longer lasting UV emission.  

\begin{figure}[t]
\center 
\includegraphics[scale=0.65,angle=0]{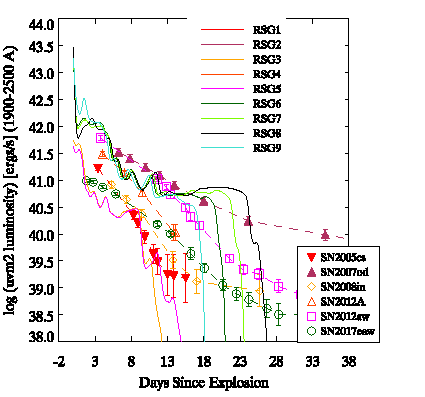}
\includegraphics[scale=0.65,angle=0]{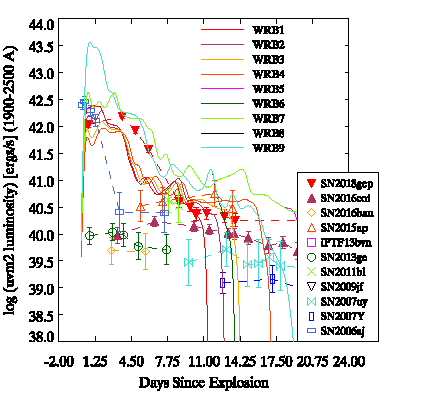}
\includegraphics[scale=0.65,angle=0]{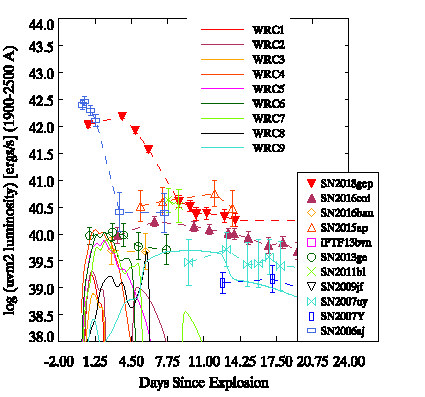}
\caption{Early-time UV fluxes as a function of time for a broad range of 
observations and models \citep{Brown_etal_2014}.  A variety of UV light-curves have been observed with peak brightness measurements that range over an order of magnitude in inferred luminosity.  Our models are designed to span this range, calibrated by these observations.}
\label{fig:UVcomp}
\end{figure}

Our LANL light-curve calculations also produce a broad set of UV light-curves and spectra.  We can use the UV emission to further calibrate our simulations using the more extensive set of early-time UV observations of SNe from the UltraViolet Optical Telescope (UVOT; \citealp{rom04}) on the Neil Gehrels Swift Observatory \citep{geh04}.  Figure~\ref{fig:UVcomp} compares a set of our models to a set of UV observation from the Swift Optical Ultraviolet Supernova Archive (SOUSA; \citealp{Brown_etal_2014}).   The observed UV emission spans a broad range of luminosities and our models also span this broad range of emission.  This UV emission guided the development of our three classes of WR models in Table~\ref{tab:Models}.  From the fitting of our models to the UV emission, we designed our normal and low distributions of X-ray light curves used in Section~\ref{sec:rate}.

\section{Understanding the Early-Time Signal}
\label{sec:sample}

Asymmetries in the explosion, the progenitor star, and the wind all produce a series of shocks that extend and complicate the early shock signal~\cite[e.g.][]{2020ApJ...898..123F,2021arXiv210913259I}.    These signals produce a range of peak emission and durations.  It will be difficult to disentangle these effects with a single supernova observation.  To study (and discriminate between) all of these effects, we need a large set of supernova observations (light-curves, hardness ratios, and spectra), but here we focus on what is needed to determine the distribution of peak luminosities and  durations.

To get a handle on the number of systems needed to constrain the properties of the shock breakout signal, we construct a 2-component model for the duration of this shock breakout emission:  a short-duration component expected from a spherical outburst with a average duration of 400 s and a long-duration component driven by asymmetries with and average duration of 7000s.  For each of these, we assume a Gaussian distribution with a standard deviation of 100 s and 1000 s for the short, long components respectively.  Using a Monte Carlo approach, we then determine the extent at which we can constrain these two distributions with a finite sampling size.

 For our samples, we assume that the time resolution of the observations are 8\,s, limiting that accuracy of any duration to that 8\,s window.  We expect that once a detection of a shock breakout event occurs, observers can improve the time resolution of the measurements and, for example, a 24\,s detection measurement can be improved to 8\,s time resolution after the detection.  For each sample size, we draw randomly from our distribution and determine the sample's estimate of the average duration of each component and its standard deviation.  We repeat this process, using multiple instances using a Monte Carlo approach for each sample size to determine the accuracy as a function of sample size.  This Monte Carlo approach is achieved in the following way:  for a given sample size $n$, we draw $n$ SN from our 2-component model.  From these $n$ supernovae, we infer the best fit 2-component distribution by using a $\chi^2$ test to determine both the average duration of the long and short component as well as the standard deviation in the timescale of these two components.  We repeat this process 100,000 times to determine the fraction of $n$ samplings that accurately reproduce the average and standard deviation of the duration to a given percentage accuracy.

\begin{figure}[t]
\center 
\includegraphics[scale=0.45,angle=0]{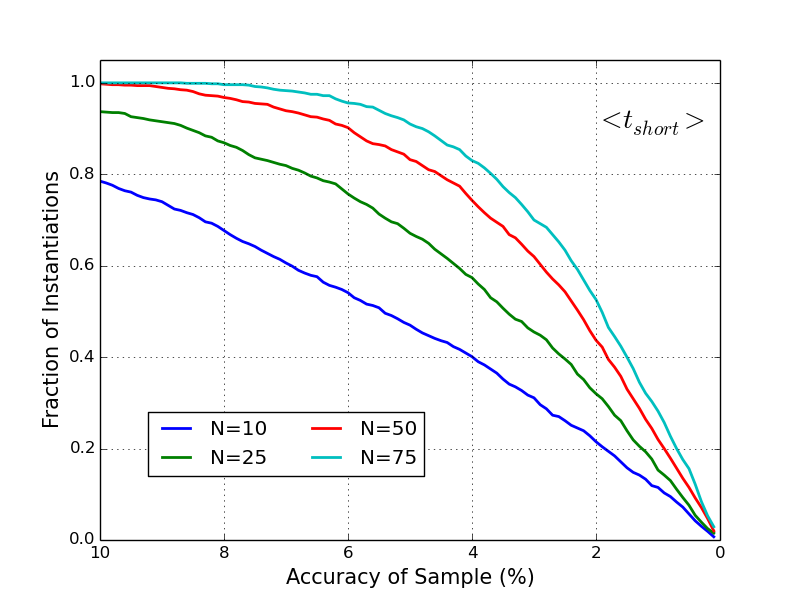}
\includegraphics[scale=0.45,angle=0]{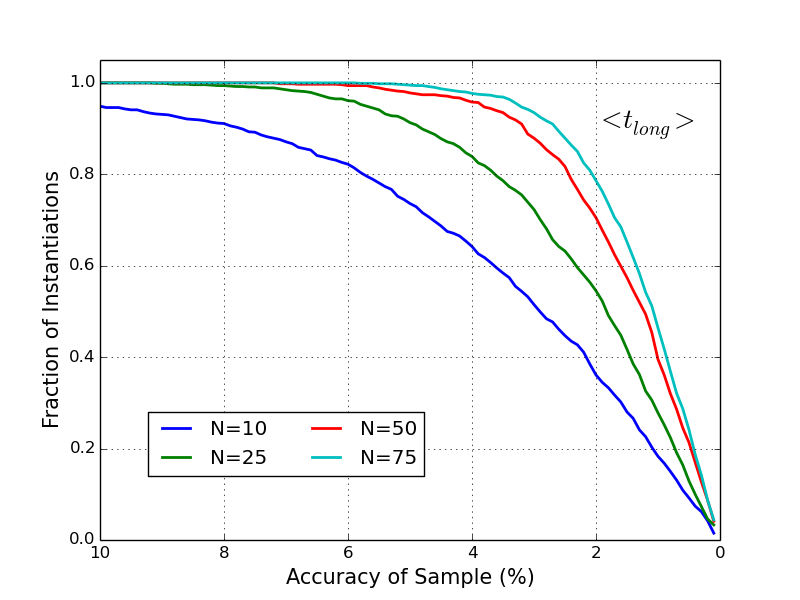}
\caption{Fraction of sampling instantiations that match the parent distribution to a given accuracy for 4 different sampling sizes for both the short and long duration components.  Because of our 8\,s window, a larger sample is required to get a high-level of accuracy in the short duration component.  See text for details on these models.}
\label{fig:sampleave}
\end{figure}

 Figure~\ref{fig:sampleave} shows the fraction of sampling instances where the average value for the duration distribution matches our parent distribution to within a a given percentage.  For example, from this figure we note that 80\% of our 100,000, $n=10$ instantiations predicted an average short duration timescale within 10\% of the true value (Fig.~\ref{fig:sampleave}) with only 40\% predicting a value within 4\% of the true value.  The 8\,s observing window means that achieving a high level of accuracy for the short-duration component requires a larger sampling size.  Alternatively, if we observe over 25 early-time emission events, we have over a 90\% chance to accurately estimate the average duration of this short-duration component to better than 10\%.  With 50 observations, this chance is raised to above 99\%.

\begin{figure}[t]
\center 
\includegraphics[scale=0.45,angle=0]{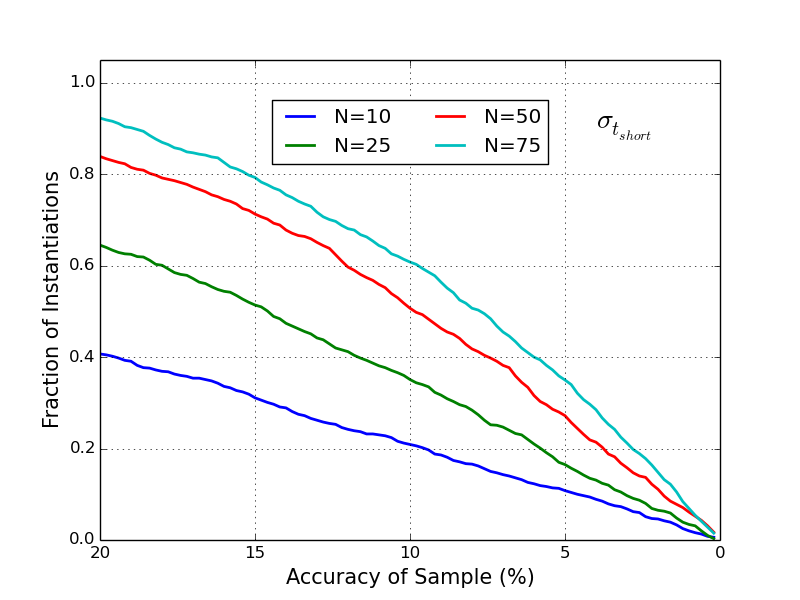}
\includegraphics[scale=0.45,angle=0]{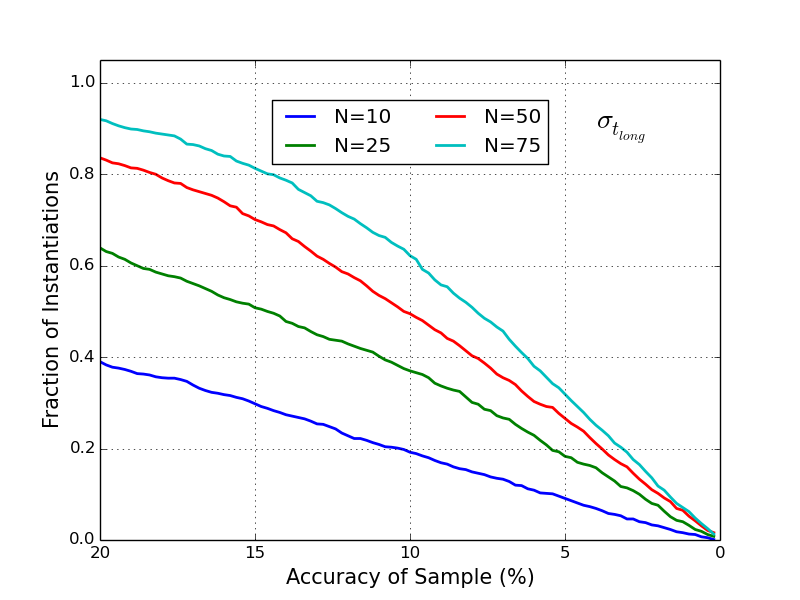}
\caption{Fraction of sampling instantiations that match the parent standard-deviation distribution to a given accuracy for 4 different sampling sizes for both the short and long duration components.  The standard deviation is much more difficult to constrain than the average value (see Figure~\ref{fig:sampleave}).}
\label{fig:samplesig}
\end{figure}

The corresponding inference of the standard deviation is shown in Figure~\ref{fig:samplesig}.  Achieving a high level of accuracy in the standard deviation is more difficult.  Even for a sample of 75 systems, we only have a 90\% chance to measure this standard deviation to 20\% for both our short- and long-duration models.

The true sample is more complex than a simple 2-Gaussian component model used here.  It is likely that the signal will be more blended with different components contributing to extended signals.  A dataset will ultimately have to be compared to detailed models including the full set of physics behind this early-time emission.  However, any observed sample will have much more data than the distribution of durations.  If we couple detailed models to luminosities, colors, and spectra to the duration measurement, we will be able to further constrain the models and the distribution of systems.  This highlights the need to not only get multiple signals, but ensure we are capture the full signal for each.   Since the signal often peaks in the X-ray, X-ray measurements, along with UV signals, are crucial.

\section{Detection Rates}
\label{sec:rate}

As we have shown in Section~\ref{sec:sample}, to use early-time shock emission to constrain stellar and explosion properties, we must observe a large sample of systems, not just a single event.
To estimate the detection rates of possible future satellite missions, we combine the achievable instrumental sensitivities, modeled SN luminosities, and SN rates in a simulation.  
\subsection{SN Rates}

We first calculate the core-collapse SNe (CCSNe) rate. The rate we use for CCSNe is $0.72 \times 10^{-4} \textrm{yr}^{-1} \textrm{Mpc}^{-3} \textrm{h}_{70}^{ -1}$, for redshift $z=0-0.04$ \citep{str15}.  Assuming a two year mission and a 200 Mpc distance limit, there are a total of 4825 CCSN in the simulation.  We used a limit of 200 Mpc as this created a good sample without the code running for a lengthy amount of time.  After being confident in our modeling, we also ran simulations out to 500 Mpc (see Table \ref{ccsn}). For the simulation the SN events are distributed in shells of width 10 Mpc according to the SN rate and volume of the shell.    The distribution by type is also based on Table 1 of \citet{str15}.    We define in our simulation red supergiants (RSG) as Type IIP and IIL and comprise 59.70\% of the total. Wolf-Rayets (WR) are Type Ib, Ic, and IIb and comprise 31.34\% of the total. Blue super giants (BSG) are inherently rare and we assign only 1\% of the total to these.  This leaves 9.76\% for ``Other" types (e.g. Type IIn and other exotics). The total by type is: 2880 RSGs, 1512 WRs, 48 BSGs, and 385 others.  

\begin{figure*}
\center 
\includegraphics[scale=0.30,angle=0]{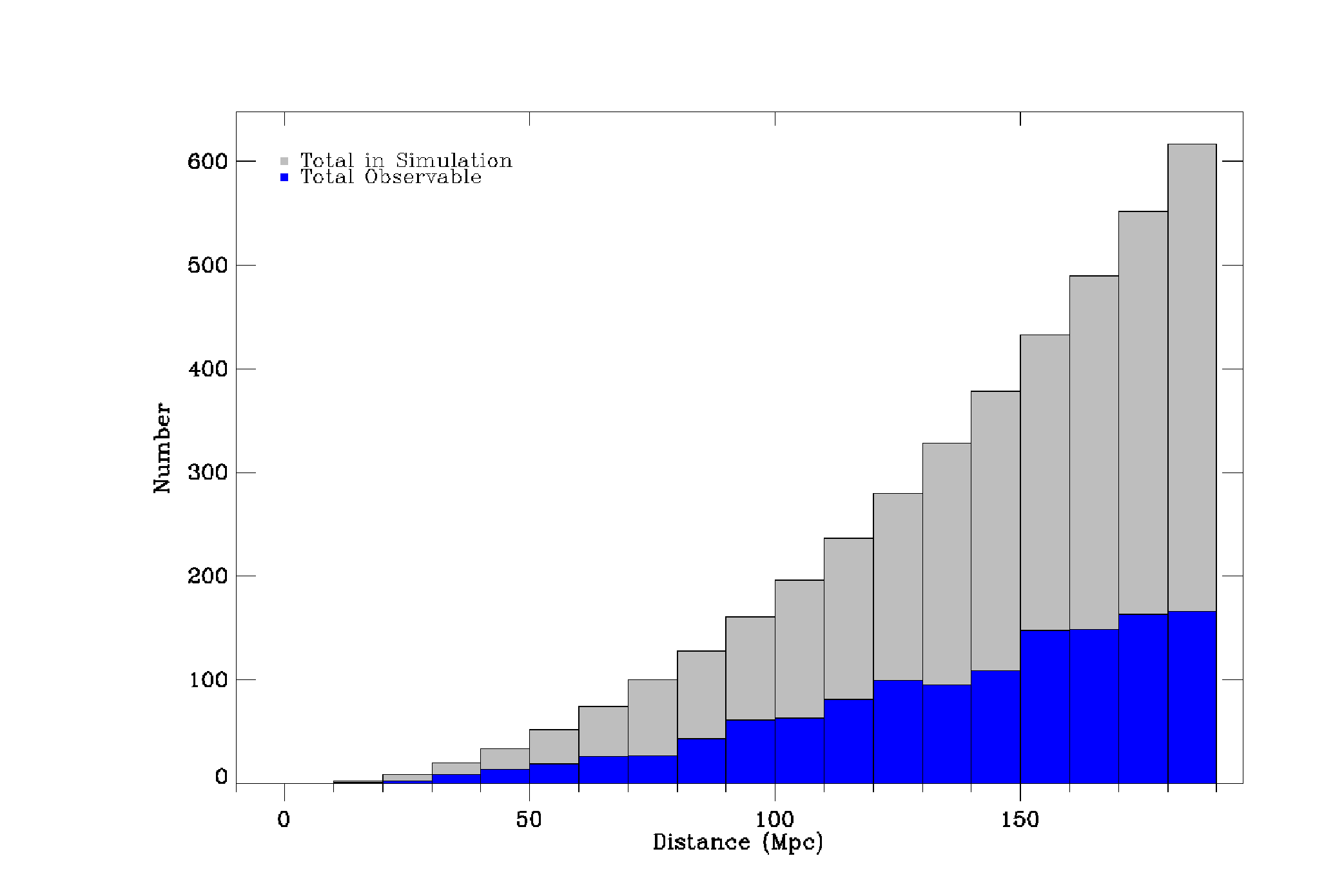}
\caption{Distribution of CCSNe produced in the simulation (gray) and available for observations (blue) per 10 Mpc bin. Available for observing means detectable by an X-ray sensor with a limiting sensitivity of $1.6\times10^{-10}$ erg/s/cm$^{2}$. We also note that the number of observable SNe falls off rapidly in simulations extended further than 200 Mpc.  To be conservative in number estimations, we use the 200 Mpc cut off. }
\label{galbin}
\end{figure*}

Next we must determine the location and distance of each event. For a galaxy database we use a nearby sample released by the Pan-STARRS survey \citep{ps1}. We extracted from this catalog galaxies that were within 200 Mpc (z $\approx$ 0.05, Figure \ref{galbin}).  As noted prior, we also ran the simulation out to 500 Mpc, but this requires a volume extrapolation of the galaxy distribution. The limit of 200 Mpc allows us to confidently select real galaxies from the Pan-STARRS catalog. This produced over six million galaxies in a 200 Mpc spherical volume.   For each SN the simulation randomly chooses a galaxy within the 10 Mpc bin, thus giving a RA, DEC, and a distance to the event.  Using the catalog is important for a mission simulation as this will account for galaxy clusters and give a more accurate placement of the SNe events over a random sky distribution. The survey information did not give galaxy type (spiral, elliptical, etc.) and we do not make any stipulation on star forming galaxy locations. This should not be a large issue for observable SNe as the vast majority of galaxies have of X-ray luminosity $Log~L_X < 41$ erg/s \citep{geo08}.  In most cases our simulations have SNe that survive to be observable have $Log~L_X > 42$ erg/s.  For the extrapolation to 500 Mpc, the galaxy distances are assigned randomly from the galactic distribution. The distribution extrapolation to 500 Mpc is estimated from the Pan-STARRS distribution and the galactic catalog of  \citet{whi11,whi12}.  %Figure\ref{galbin} shows the number of galaxies in the simulation per 10 Mpc bin. 

\begin{figure}[htb]
\center 
\includegraphics[scale=0.40,angle=90]{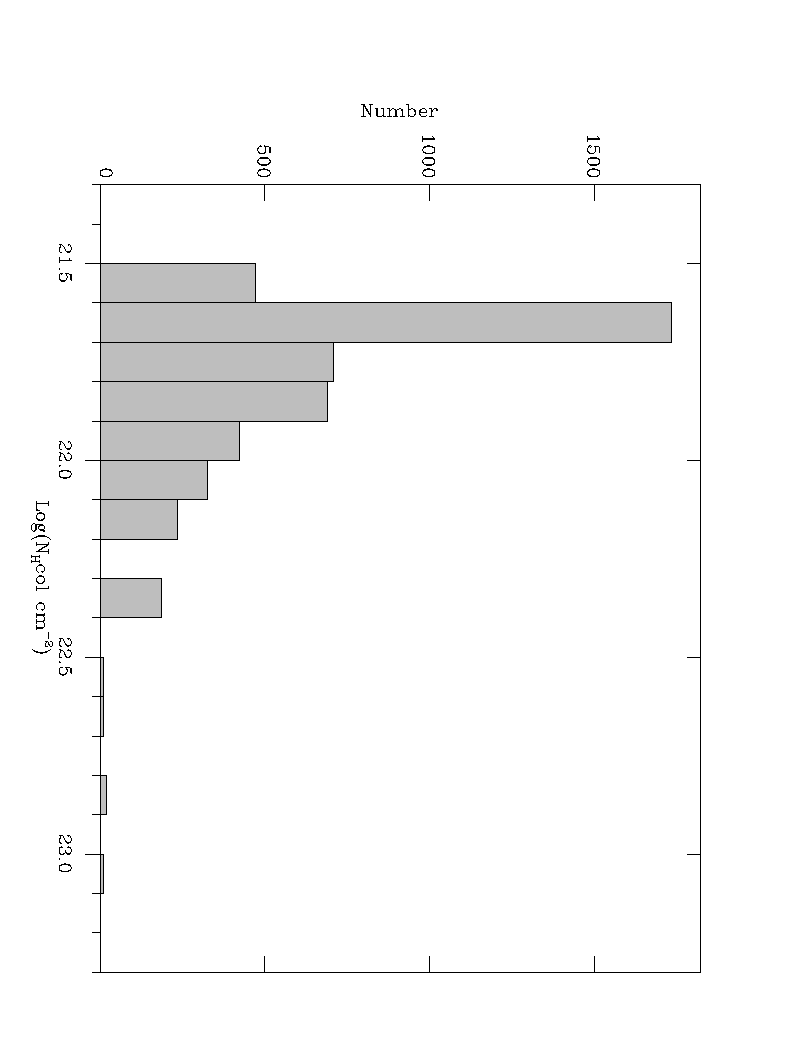}
\caption{Distribution of column density in the simulation.}
\label{ncolhist}
\end{figure}

To determine the observability of the X-ray emission, we use the two sets of models for RSGs, BSGs, and WR stars described above, namely the "Gaussian" model based on \citet{nak10} and a "LANL" model described above.
%:  one using X-ray observations to calibrate analytic models and one using UV observations to constrain radiation-transport models of parameterized SN explosions (see Section~\ref{sec:LCM}).  
With these two sets of models, we place the SNe at the distance of its assigned galaxy and assign a column density for the extinction.  To assign a column density to each SNe, we assume a Galactic column density of $N_{col}=3.6\times10^{21} cm^{-2}$ \citep{sch12}. 
We then add a random host galaxy $N_{col}$ based on the distribution in Figure 1 in \citet{wil13}.  The distribution of hydrogen column density is shown in Figure \ref{ncolhist}.   We then determine if the SNe would be visible for the given brightness limit of our assumed detector.   If the simulated light curve is above the threshold of detection for a 24 second X-ray integration, the SN is flagged as one that could be observed in the X-ray if the telescope is scanning the region of the sky it is occurring in.  The duration of observability is the time the brightness is above this threshold.
This is important for a mission as even if the initial shock emergence is missed, this will determine how many young SNe will be seen as the spacecraft slews to fields in which a young SN has recently exploded.  

\subsection{Determining UV Visibility}
The next step is to determine if the X-ray observable candidates could also be observed in the UV by the nominal UV instrument defined above.
Each of the simulated SN explosions above has an associated UV luminosity.
Just as for the X-ray light curves, we determine the brightness by attenuating by the simulated distance to the event.  Using the assigned column density, we calculate for each SNe $E(B-V) (mag) = n_{H} / 5.3 \times 10^{21} {\rm cm^{-2}}$ \citep{pre95}. This is converted to the UV extinction at 1900 \AA\ using the extinction law of \citet{car89}.   Arising from young, massive progenitors, CCSNe usually explode in regions of their host galaxy with UV emission from their star forming regions.  To quantitatively measure the impact of this emission, we measured the UV luminosity from UVOT uvm2 images of the same CCSNe shown in Figure \ref{fig:UVcomp}.  The UV luminosities ranged from log(L$_{uvm2})\sim39-41$.  These host galaxy luminosities can be brighter than the fainter WRC models, but are an order of magnitude fainter than the RSG and WRB models which are observable in our simulation. We adopt an integrated observation flux threshold of $8.0 \times 10^{-14}$ erg/s/cm$^{2}$ in the UV.  We determine the ability of the UV light curve to be observed the same way as the X-ray light curve.

 The number of SNe visible all sky within 500 Mpc and 200 Mpc over two years is in Table \ref{ccsn} for a 24 second exposure and the sensitvity given in Table \ref{tab:Instrument}. We also tested a low luminosity (Gaussian and LANL) RSG model with peak luminosity of $Log (L_{\rm X-ray}) = 42~ {\rm  erg s^{-1}}$ and WR model with peak luminosity of $Log (L_{\rm X-ray}) = 43~ {\rm  erg s^{-1}}$.  The peak luminosity for the BSG and Others were the same.  All the distributions had a 3-sigma Gaussian average spread in luminosity instead of 1-sigma. These results are also Table \ref{ccsn}. The wider peak luminosity distribution on the RSG allows for brighter SNe, even though average luminosity is fainter overall.  The values in Table \ref{ccsn}  should be regarded as the range between optimistic and conservative numbers for a potential mission.

\begin{table*}[htb]
\footnotesize
\center
\caption{Number of CCSNe Visible All-Sky Assuming a Limiting X-Ray Sensitivity of $1.2\times10^{-10}$ erg/s/cm$^{2}$.  This shows a comparison between a higher peak luminosity average with a narrower (1-sigma) distribution versus a lower luminosity peak with a wider (3-sigma) distribution. The values here represent an average from five simulations with the standard deviation from simulation to simulation noted. }
\label{ccsn}
\begin{tabular}{lcccccccc}
\hline
&Gaussian& Gaussian& Gaussian&Gaussian&LANL& LANL &LANL& LANL \\
&1-sigma&1-sigma&3-sigma&3-sigma&1-sigma&1-sigma&3-sigma&3-sigma\\
&X-ray&UV&X-ray&UV&X-ray&UV&X-ray&UV\\
\hline
Total SNe at 500 Mpc:&	8232$\pm$37 & 5404$\pm$56 & 12567$\pm$134 & 8831$\pm$77 & 9683$\pm$61 & 6587$\pm$46 & 13028$\pm$107 & 10767$\pm$111\\

Total SNe at 200 Mpc:& 1137$\pm$18	& 760$\pm$18 & 1125$\pm$37 & 818$\pm$20 & 1276$\pm$23 & 942$\pm$21 & 1178$\pm$17 & 998$\pm$12\\
\hline
\hline
By Type: &&&&&&\\
RSG (500 Mpc):	&	668$\pm$33 & 324$\pm$20 & 5827$\pm23$ & 3881$\pm$41 & 655$\pm$22 & 563$\pm$21 & 5832$\pm$62 & 5128$\pm$68\\
RSG (200 Mpc):	&	209$\pm$15 & 111$\pm$13 & 540$\pm$21 & 368$\pm$21 & 197$\pm3$ & 168$\pm$9 & 556$\pm$13 & 496$\pm$17\\

BSG (500 Mpc):&		71$\pm$12& 36$\pm$7 & 224$\pm$14 & 163$\pm$9 & 72$\pm$7 & 58$\pm$8 & 246$\pm$10 & 210$\pm$15\\
BSG (200 Mpc):&		13$\pm$4 &	6$\pm$2 & 18$\pm$2 & 15$\pm$3 & 11$\pm$3 & 10$\pm$3 & 19$\pm$3 & 19$\pm$3\\
							
WR (500 Mpc):&		6905$\pm$22 & 4742$\pm$49 & 4592$\pm$123 & 3408$\pm$75 & 8366$\pm$37 & 5450$\pm$46 & 5041$\pm$70 & 3727$\pm$59\\
WR (200 Mpc):&		816$\pm$16 & 590$\pm$15 & 402$\pm$11 & 309$\pm$15 & 960$\pm$15 & 666$\pm$13 & 445$\pm$19 & 336$\pm$14\\
									
Other (500 Mpc):	&588$\pm$12 & 303$\pm$10 & 1925$\pm$22 & 1379$\pm$29 & 591$\pm$17 & 515$\pm$17 &1909$\pm$46 & 1702$\pm$43\\
Other (200 Mpc):	&100$\pm$8 & 53$\pm$11 & 165$\pm$7 & 125$\pm$10 & 109$\pm$9 & 98$\pm$8 & 159$\pm$11 & 147$\pm$9\\

\end{tabular}
\end{table*}

\section{Conclusion} 

Shock breakout has the potential to provide one of the most direct probes of supernova explosions and their progenitors.  But to use these as probes, we need to move from observing 1 or 2 serendipitous shock breakout events to observing a large sample of shock breakout signatures.  In this paper, we designed a suite of shock breakout signatures using both analytic and simulated light-curves.  We demonstrated that, to use this early time emission to constrain properties of the explosion and its progenitor,  we must understand both the distribution of peak luminosities and the signal durations.  This, in turn, requires a large sample ($>25$) of observed events.  With the different light-curve properties and using the latest results for galaxy distributions and star formation in the nearby universe, we estimate the number of shock breakout events that could be observed with current detector technology.

The science impact of observing the early emission from shock breakout events cannot be understated. Proposed missions, such as the Astrophysical Transients Observatory (ATO; Roming et al., 2018), 
%and the Shock Interaction and Breakout Explorer (SIBEX; Roming et al., 2021), 
would provide the necessary tools for rapidly observing these shock break out events.  Our simulations show that such a mission as ATO, with full sky coverage, could detect over 1000 events.  Even with sky coverage of a fraction of the sky ($\approx0.2$ sr), a wide-field detector could increase the number of detected shock-breakout events by 2 orders of magnitude.  Because the detection is not planned for serendipitous events, the current observations do not have full coverage of the event.  A systematic study would allow us to dramatically increase our understanding of shock breakout and its constraints on supernova explosions and their progenitors.

\begin{acknowledgements}

The work by CLF was supported by the US Department of Energy through the Los Alamos National Laboratory. Los Alamos National Laboratory is operated by Triad National Security, LLC, for the National Nuclear Security Administration of U.S.\ Department of Energy (Contract No.\ 89233218CNA000001).  PJB's work on the UV properties of core-collapse SNe is supported by NNX17AF43G.

\end{acknowledgements}

%% For this sample we use BibTeX plus aasjournals.bst to generate the
%% the bibliography. The sample631.bib file was populated from ADS. To
%% get the citations to show in the compiled file do the following:
%%
%% pdflatex sample631.tex
%% bibtext sample631
%% pdflatex sample631.tex
%% pdflatex sample631.tex

\bibliography{SNeSim}{}
\bibliographystyle{aasjournal}

%% This command is needed to show the entire author+affiliation list when
%% the collaboration and author truncation commands are used.  It has to
%% go at the end of the manuscript.
%\allauthors

%% Include this line if you are using the \added, \replaced, \deleted
%% commands to see a summary list of all changes at the end of the article.
%\listofchanges

\end{document}